\documentclass[prd,aps,twocolumn,showpacs,amsmath,amssymb,superscriptaddress,floatfix,nofootinbib,9pt]{revtex4}
\usepackage[dvipdfm,colorlinks=true, citecolor=blue, linkcolor=blue, urlcolor=blue]{hyperref}
\usepackage{xcolor,multirow,amsfonts,graphicx,amsfonts,multirow,amsmath}
\newcommand{\ti}{\title}
\newcommand{\aaa}{\author}
\newcommand{\ee}{\email}
\newcommand{\add}{\address}

\newcommand{\DS}[1]{/\!\!\!#1}
\newcommand\grap[2]{\includegraphics[width=#1\textwidth]{#2}}

\newcommand{\bdoc}{\begin{document}}
\newcommand{\edoc}{\end{document}}

\newcommand{\babs}{\begin{abstract}}
\newcommand{\eabs}{\end{abstract}}

\newcommand{\bbib}{\begin{thebibliography}{99}}
\newcommand{\ebib}{\end{thebibliography}}

\newcommand{\bwt}{\begin{widetext}}
\newcommand{\ewt}{\end{widetext}}

\newcommand{\beq}{\begin{eqnarray}}
\newcommand{\eeq}{\end{eqnarray}}

\newcommand{\bali}{\begin{align}}
\newcommand{\eali}{\end{align}}

\newcommand{\bit}{\begin{itemize}}
\newcommand{\eit}{\end{itemize}}

\newcommand{\bfig}{\begin{figure}}
\newcommand{\efig}{\end{figure}}

\newcommand{\bfigs}{\begin{figure*}}
\newcommand{\efigs}{\end{figure*}}

\newcommand{\btab}{\begin{table}}
\newcommand{\etab}{\end{table}}

\newcommand{\btabu}{\begin{tabular}}
\newcommand{\etabu}{\end{tabular}}

\newcommand{\btabs}{\begin{table*}}
\newcommand{\etabs}{\end{table*}}

\newcommand{\SecI}{\section{Introduction}\label{section:1}}
\newcommand{\SecII}{\section{Calculation Technology}\label{section:2}}
\newcommand{\SecIII}{\section{Numerical Results}\label{section:3}}
\newcommand{\SecIV}{\section{Summary}\label{section:4}}

\newcommand{\OHn}{\langle {{\mathcal O^H}(n)}\rangle}
\newcommand\nSn[2]{[^#1S_#2]}

\newcommand\bQ[1]{\langle b#1\rangle}
\newcommand\Xc{\Xi_{bc}}
\newcommand\Xb{\Xi_{bb}}
\newcommand\XQ{\Xi_{bQ'}}
\newcommand\g{\gamma}
\newcommand\G{\Gamma}
\newcommand\B{{\cal B}}
\newcommand\R{{\cal R}}
\newcommand\nn{\nonumber}
\newcommand\ii{\item}
\newcommand\dd{\delta}
\newcommand\Cijk{C_{ij,k}}
\newcommand\Gmnk{G_{mnk}}
\newcommand\Ta[1]{ (T^a)_{#1}}

\allowdisplaybreaks
\date{\today}

\bdoc
\ti{Investigation for $Z$-boson decay into $\Xc$ and $\Xb$ baryon with the NRQCD factorizations approach}
\aaa{Xuan Luo}
\ee{cnluoxuan@hotmail.com}
\aaa{Hai-Bing Fu}
\ee{fuhb@gzmu.edu.cn}
\aaa{Hai-Jiang Tian}
\add{Department of Physics, Guizhou Minzu University, Guiyang 550025, P.R. China}

\babs
The $Z$-boson decay provides good opportunities for the research on $\XQ$ baryon due to large quantity of $Z$ events that can be collected at the high-energy colliders. We performed a completed investigation of the indirect production of the $\Xc$ and $\Xb$ baryon via $Z$-boson decay $Z\to \XQ+\bar b +\bar Q'$ with $Q'= (c,b)$ quark according to NRQCD factorizations approach. After considering the contribution of the diquark states  $\bQ{c}\nSn{3}{1}_{\bar 3/6}$, $\bQ{c}\nSn{1}{0}_{\bar 3/6}$, $\bQ{b}\nSn{1}{0}_{6}$ and $\bQ{b}\nSn{3}{1}_{\bar 3}$, the calculated branching ratio for $Z\to\XQ+X$ are $\B(Z\to\Xc+X) = 3.595\times 10^{-5}$ and $\B(Z\to\Xb+X) = 1.213\times 10^{-6}$. Moreover, the $\Xc$ events produced are predicted to be of the $10^4(10^7)$ order at the LHC(CEPC), while the $\Xb$ events produced are forecasted to be of the $10^3(10^6)$ order. Furthermore, we have estimated the production ratio $\R(Z_Q\to\Xi^{+,0}_{bc})$ with four $Z$-boson decay channels. The $\R(Z_Q\to\Xi^{+,0}_{bc})$ up to $10^{-6}$ for $Z\to c\bar c$ channel and $10^{-5}$ for $Z\to b\bar b$ channel, respectively. Finally, we present the differential decay widths of $\Xc(\Xb)$ with respect to $s_{23}$ and $z$ distributions, and analysis the uncertainties.
\eabs

\date{\today}

\maketitle

	  \SecI

Doubly heavy baryons consisted with two heavy quarks and one light quark are expected within the quark model~\cite{Gell-Mann:1964ewy, Ebert:1996ec, Gerasyuta:1999pc, Itoh:2000um}. The investigate of the doubly heavy baryons is enthralling as it provides unique test for the perturbative Quantum Chromodynamics (pQCD) and the nonrelativistic QCD (NRQCD). In the past decades, research on the doubly heavy baryons related studies has developed rapidly, including both experimental and theoretical aspects.

From the experimental side, the $\Xi_{cc}^{++}$ baryon was firstly observed by the LHCb collaboration through the decay channel $\Xi_{cc}^{++} \to \Lambda _c^+ K^- \pi^+ \pi^- $ and $\Lambda_c^+ \to p K^- \pi^+$ in 2017~\cite{LHCb:2017iph}, which was identified by Ref.~\cite{LHCb:2019qed} and also by Ref.~\cite{LHCb:2018pcs} via measuring diffirent decay channel $\Xi_{cc}^{++} \to \Xi_c^+\pi^+$ with $\Xi_c^+ \to p K^- \pi^+$. Moreover, the observations of doubly charmed baryon $\Xi_{cc}^+$ was firstly reported in $\Xi_{cc}^+\to pD^+K^-$ decay channel by the SELEX collaboration. The SELEX collaboration announced observations of production rates of $\Xi_{cc}^+$, which were not confirmed by the FOCUS~\cite{Ratti:2003ez}, BABAR~\cite{BaBar:2006bab}, and Belle~\cite{Belle:2006edu}, where the collision energy of FOCUS is comparable with SELEX. Over the past few years, the LHCb collaboration has published their observation of $\R(\Xi^+_{cc})$, defined as $\R(\Xi^+_{cc}) = \sigma(\Xi_{cc}^+) \B(\Xi_{cc}^+\to\Lambda_c ^+ K^- \pi^+)/\sigma(\Lambda_c^+)$~\cite{LHCb:2019gqy}, varying in the region $[0.9,6.5]\times 10^{-3}$ for $\sqrt s = 8~{\rm TeV}$, and $[0.12, 0.45]\times 10^{-3}$ for $\sqrt s =13~{\rm TeV}$, these values are still significantly lower than the $\R(\Xi^+_{cc})= 9\%$ measured by the SELEX Collaboration. As regards $\Xc$, which is containing one bottom quark and one charm quark. Due to its unique nature in the family of baryons, $\Xc$ baryon also attracts widely attention of experiment and theory. In 2020, the LHCb Collaboration seek for the doubly heavy $\Xc^0$ baryon via its decay to the $D^0 p K^- $, but no evidence was found~\cite{LHCb:2020iko}. Recently, $\Xi^0_{cb}$ and $\Omega^0_{cb}$ are detected via $\Lambda_c^+\pi^-$ and $\Xi_c^{+}\pi^-$ decay modes, but evidence of signal is not found~\cite{LHCb:2021xba}. $\Xb$ is still not experimentally detected. In a nutshell, there is still no solid signal of the $\XQ$ baryon with $Q'$ is $c$ or $b$ quark. In order to investigate the baryon production properties and further testing of the NRQCD, there are lots of works has been done ~\cite{Brodsky:2017ntu, Kiselev:1994pu, Falk:1993gb, Chang:2006xp, Baranov:1995rc, Bodwin:1994jh, Gunter:2001qy,Kiselev:1995xe, Berezhnoy:2006mz,Braguta:2002qu, Braaten:2003vy,
Li:2007vy, Yang:2007ep, Bi:2017nzv, Zhang:2011hi, Jiang:2012jt,Jiang:2013ej,Martynenko:2013eoa,Yang:2014tca,Yang:2014ita,Martynenko:2014ola,Lai:2014iji,
Koshkarev:2016rci,Koshkarev:2016acq,Groote:2017szb,Yao:2018zze,Chang:2006eu,Chen:2014hqa,Zheng:2015ixa,Chen:2018koh,Berezhnoy:2018krl,Chen:2019ykv, Wu:2019gta,Niu:2018ycb,Zhang:2022jst,Niu:2019xuq,Luo:2022jxq}, both direct and indirect production.

\bfigs[t]
\grap{0.98}{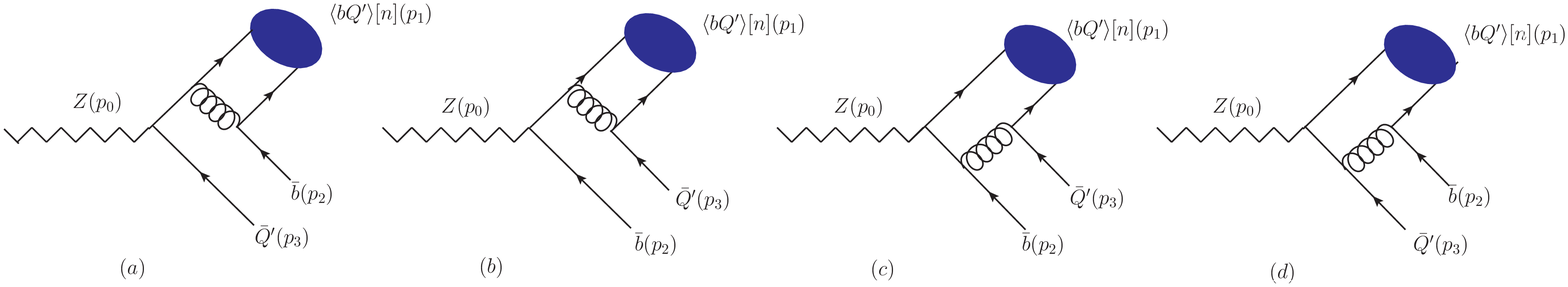}
\caption{The diagrams for $Z \to \bQ{Q'} [n] +\bar b+\bar Q'$ at Leading Order, here $Q'$ represent $c$-quark for the $\Xc$, and $Q'$ represent $b$-quark for the $\Xb$.}
\label{fig:1}
\efigs

In comparing with the direct production such as hadroproductions, photoproductions and the $e^+e^-$ annihilation, the indirect production is also fascinating, which can be attributed to the character of baryon and the properties of its initial particles. Production of $\XQ$ baryons via the top quark decays was discussed in Ref.~\cite{Niu:2018ycb}, and via $W^+$-boson decay was calculated in Ref.~\cite{Zhang:2022jst}. The $H\to \Xi_{QQ'}+X$ process was calculated in Ref.~\cite{Niu:2019xuq}. Apart from these process, the baryon can be production in $Z$ decay. Recently, the process $Z\to \Xi_{cc}+X$ have been finished~\cite{Luo:2022jxq}. The authors there found about $10^4(10^7)$ $\Xi_{cc}$ events can be detected via $Z$ decay at the LHC(CEPC) peer year, and the branching ratio $\B(Z\to\Xi_{cc}+X)$ is comparable with the $\B(Z\to J/\psi+X)$,  representing  its experimental observability in $Z$ decay. Apart from the $\Xi_{cc}$, the $Z$-boson decay can also provide a good opportunity for the studies on $\XQ$ baryon due to the large quantity of $Z$ events at the high energy colliders, e.g., about $\sim 10^9$ $Z$ events can produce at the LHC each year~\cite{Liao:2015vqa},  The proposed future $e^+e^-$ collider, CEPC~\cite{CEPCStudyGroup:2018ghi}, the number of $Z$ production events is going to be $\sim 10^{12}/$year. Moreover, the authors suggest that the decay channel $\Xi^{+,0}_{bc}\to \Xi^{++}_{cc}+X$ has multiple advantage compared to $\Xi^0_{bc}\to \Lambda^+_c\pi^-$ in Ref.~\cite{Qin:2021wyh}, which will offer a direction for the experiment to detected $\Xc$. Thus, in this paper, we shall first focus our attention on the indirect production of $\XQ$ via $Z$-boson decay and further to revealed whether a considerable amount of $\XQ$ can be collected by $Z$ decay. In addition, we will first forecast $\R(Z_Q\to\Xi^{+,0}_{bc})$ in $Z$-boson decay by decaying the channnel $\Xi^{+,0}_{bc}\to\Xi^{++}_{cc}+X$.

The rest of the paper is shown as below: We demonstrate the detailed treatment in Sec.~\ref{section:2}. The phenomenological results and analyses are given by Sec.~\ref{section:3}. A brief summary are given in Sec.~\ref{section:4}.

	  \SecII

The production for the doubly heavy baryons can be formulated into two programs~\cite{Chang:2006xp,Ma:2003zk,Chang:2006eu,Wu:2019gta}: 1) The first step is that a binding state is produced, namely diquark $\langle QQ'\rangle[n]$, where $[n]$ represent the color- and spin- combinations. Bases on the decomposition $3\otimes 3= \bar{\bf 3}\oplus \mathbf 6$ in $SU_c(3)$ group and NRQCD~\cite{Petrelli:1997ge,Bodwin:1994jh}, the quantum number of color is only the color-antitriplet and the color-sextuplet, denoted as $\bar{\bf 3}$ and $\bf 6$, respectively. And the quantum counts of the diquark $\langle QQ'\rangle$ state can be $[^3S_1]$ or $[^1S_0]$. 2) The next procedure is that the diquark fragments to a observable baryon $\Xi_{QQ'q}$ by hunting a light quark from `environment' with a fragmentation probability of almost one hundred percent. For convenience, throughout the paper, we utilize the label $\Xi_{QQ'}$ instead of $\Xi_{QQ'q}$. Among this total ``$100 \%$'' fragmentation probability, the probability of both $\Xi_{QQ'd}$ and $\Xi_{QQ'u}$ is $43\%$, respectively, and the ratio for $\Omega_{QQ's}$ is $14\%$~\cite{Sun:2020mvl,Chen:2018koh}.

The diagrams for $Z(p_0) \to \bQ{Q'}[n](p_1) +\bar b(p_2)+\bar Q'(p_3)$ at tree level are shown in Fig.~\ref{fig:1}. Then, one can be obtained the differential decay width of the process $Z (p_0) \to \bQ{Q'}[n]({p_1}) +\bar b(p_2)+\bar Q'(p_3)$ by using NRQCD factorization program~\cite{Bodwin:1996tg,Petrelli:1997ge}, which as follows:
\begin{align}\label{eq:1}
d\G=&\sum_{n}d\hat\G(Z \to \bQ{Q'}[n] +\bar b+\bar Q')\OHn
\end{align}
The $\OHn$ is symbol of long-distance matrix element, which stand for the hadronization of the diquark state $\bQ{Q'}[n] $ into the
observable baryon state $\XQ$. Genarally, $\OHn$ could be derived from the origin value of the Schr\"{o}dinger wavefunction In this paper, $\OHn = (|\Psi_{bQ'}(0)|,|\Psi^{\prime}_{bQ'}(0)|)$ for $S$-wave and $P$-wave, which are derived from the experiment data or some non-perturbative theoretical methods, e.g. the potential model, lattice QCD and QCD sum rules~\cite{Bagan:1994dy, Kiselev:1999sc, Bodwin:1996tg}.

Then, the differential decay width $d\hat\G(Z \to \bQ{Q'}[n] +\bar b+\bar Q')$ have the following form
\beq
d\hat\G(Z \to \bQ{Q'}[n] +\bar b+\bar Q')=\frac13 \frac1{2m_z}{\sum |M[n]|^2}d\Phi_3,
\label{eq:2}
\eeq
with $m_z$ indicate the $Z$-boson mass , $|M[n]|$ being the hard amplitude expressions, the constant $1/3$ was given by the spin average of the initial $Z$-boson, and $\sum$ means that we need to sum over the color and spin of all the final particles. The three-body phase space $d\Phi_3$ follow as
\beq
d\Phi_3 = (2\pi)^4 \dd^4 \bigg(p_0 - \sum\limits_f^3 p_f\bigg) \prod \limits_f^3 \frac{d^3p_f}{(2\pi)^3 2p_f^0}.
\label{eq:3}
\eeq
The three-particle phase space with massive quark or antiquark in the final state can be found in Refs.~\cite{Chang:2007si,Wu:2008cn}. Then, the decay width can be rewritten as
\beq
&&d\hat\G(Z\to\bQ{Q'}[n] + \bar b + \bar Q')
\nn\\
&&\hspace{2cm} = \frac1{2^8\pi^3 m_z^3} \sum |M[n]|^2 ds_{12} ds_{23}.
\label{eq:4}
\eeq
with the $s_{ij} = (p_i + p_j)^2$. For the production of the $\XQ$ baryon, diagrams for the $Z\to\bQ{Q'}[n]+\bar b+\bar Q'$ at Leading Order (LO) are listed in Fig.~\ref{fig:1}, where $Q'$ represent $c$-quark for the $\Xc$, and $Q'$ denote $b$-quark for the $\Xb$. We utilize the charge parity $C=-i\g^2\g^0$, hard amplitude expressions $M[n]$ for the production baryon will be easily obtained from the familiar meson production, which has been sufficiently documented in Refs.~\cite{Jiang:2012jt,Zheng:2015ixa}, here we have a brief descriptions.

Firstly, we use the $C=-i\g^2\g^0$ to reverse one fermion line. Generally, the fermion line which need to be reversed can be writing as $L_1=\bar u_{s_1}(k_{12}){\G_{i+1}} S_F(q_i,m_i) \cdots S_F(q_1,m_1)\G_1 v_{s_2}(k_2)$. In which $\G_i$ are the interaction vertex, the symbol for fermion propagator denote $S_F (q_i,m_i)$, $s_1$ or $s_2$ is for spin index, and $(i = 0, 1, ...)$ represent the quantity of the interaction vertices in this fermion line. According to he charge parity $C=-i\g^2\g^0$, we have
\begin{align}
&v_{s_2}^{\rm T}(p)C=-\bar \mu_{s_2}(p),&& C^{-1}\G_i^{\rm T} C=-\G_i,
\nn\\
&CC^{-1}=I,&&C^{-1}S_f^{\rm T} (-q_i,m_i)C=S_f(q_i,m_i),
\nn\\
&C^{-1}\bar\mu_{s_1}^{\rm T} (p_{12})=\nu_{s_1}(p_{12}), && C^{-1}(\g^{u})^{\rm T} C=-\g^{u},
\nn\\
&C^{-1}(\g^{u}\g^{5})^{\rm T} C=\g^{u}\g^{5}.
\label{eq:7}
\end{align}
If the fermion line does not includes axial vector vertex, we can be readily obtained the following

\beq
&&\hspace{-0.5cm}L_1 = L_1^{\rm T} = v_{{s_2}}^T(p_2)\G_1^{\rm T} S_F^{\rm T} (q_1,m_1) \cdots S_F^{\rm T} (q_1,m_1)\G_{i+1}^{\rm T} \bar u _{s_1}^{\rm T}({p_{12}})
\nn\\
&&= v_{s_2}^{\rm T} (p_2)C C^{-1}\G_1^{\rm T} C C^{-1} S_F^{\rm T} (q_1,m_1)C C^{-1}
\nn\\
&&\quad\cdots C{C^{ - 1}}S_F^{\rm T} (q_1,m_1)C C^{-1}\G_{i+1}^{\rm T} C C^{-1}\overline u_{s_1}^{\rm T} (p_{12})
\nn\\
&&= (-1)^{i+1}\bar u_{s_2}(p_2)\G_1 S_F(-q_1,m_1)
\nn\\
&&\quad\cdots S_F(-q_i,m_i){\G_{i+1}}{v_{s_1}}({p_{12}}).
\label{eq:5}
\eeq
Otherwise, through reversing the fermion line, we can obtain the amplitude of the baryon production from familiar meson production except an additional $(-1)^{(n+1)}$ coefficient for pure vector case and $(-1)^{(n+2)}$ factor for including an axial vector case. i.e. the amplitude of $Z\to \bQ{Q'}[n]+\bar b+\bar Q'$ can be written as
\beq
M_{\rm diquark}=(M^a_1-M^v_1)+(M^a_2-M^v_2)+M_3+M_4,
\eeq
with $M_i(i=1,2,3,4)$ is the hard amplitude of the familiar meson production, $M^a_i$ and $M^v_i$ denote the parts of the axial vector amplitude and the pure vector amplitude of $M_i$, respectively.

Then, the amplitude $M_l[n]$ with $l=(a, . . ., d)$ are obtained from Fig.~\ref{fig:1} according to Feynman rules, which can be read off:
\bwt
\beq
&&M_a[n]= - \kappa \frac{\bar u(p_{12})(- i\g^\nu) v (p_2)\bar u (p_{11}) (- i\g^\nu) (m_{Q'} + \DS p_1 + \DS p_2 ) \DS \epsilon(p_0) (c_v + c_a\g^5) v (p_3)}{(p_{12} + p_2)^2 \left[(p_1 + p_2)^2 - m_{Q'}^2 \right]},
\nn\\
&&M_b[n]= - \kappa \frac{\bar u (p_{12}) (- i \g^\nu ) ( m_b + \DS p_1 + \DS p_3)\DS\epsilon(p_0)(c_v + c_a\g^5) v (p_2)\bar u( p_{11}) (- i \g^\nu)v (p_3)}{(p_{11} + p_3)^2 \Big[ ( p_1 + p_3 )^2 - m_b^2\Big]},
\nn\\
&&M_c[n]= - \kappa \frac{\bar u (p_{12} )\DS \epsilon(p_0) (c_v + c_a\g^5) (m_b - \DS p_{11} - \DS p_2 - \DS p_3 ) ( - i \g ^\nu )v (p_2)\bar u (p_{11}) (- i \g ^\nu)v (p_3)} {(p_{11} + p_3)^2 \Big[ (p_{11} + p_2 + p_3 )^2 - m_b^2\Big]},
\nn\\
&&M_d[n]= - \kappa \frac{\bar u (p_{12}) ( - i \g^\nu )v(p_2) \bar u(p_{11}) \DS \epsilon(p_0) (c_v + c_a\g^5)(m_{Q'} - \DS p_{12} - \DS p_2 - \DS p_3)(-i\g^\nu)v(p_3)} {(p_{12}+ p_2 )^2 \Big[(p_{12} + p_2 + p_3 )^2 - m_{Q'}^2\Big]},
\label{eq:3x}
\eeq
where $p_{11}$ and $p_{12}$ represent the momenta of bottom quark and heavy $Q'$ quark with $Q = (c,b)$ for $\Xc(\Xb)$ production. And $\kappa=-C g_s^2$, $C$ indicate the color factor $C_{ij,k}$, and $c_{v}$,~$c_a$ are vector and axial coupling constants of $Z_{Q'\bar Q'}$ vertex. If $Q'$ representing the heavy $c$-quark, we have
\beq
c_v =- \frac{e(8\sin^2\theta_w-3)}{12\cos\theta_w \sin\theta_w},~~c_{a} =- \frac{e}{4\cos\theta_w \sin\theta_w}.
\eeq
and $Q'$ denote the heavy $b$-quark, we can obtain
\beq
c_v = \frac{e(4\sin^2\theta_w-3)}{12\cos\theta_w \sin\theta_w},~~c_{a} = \frac{e}{4\cos\theta_w \sin\theta_w}.
\eeq
Here $\theta_w$ is the Weinberg angle. With the help of Eq.~\eqref{eq:5} and insert the spin projector $\Pi_{p_1}^{[n]}$, the amplitudes can be rewritten as
\beq
&M_a[n]&= - \kappa \frac{\bar u(p_2)(-i\g^\nu )\Pi_{p_1}^{[n]} (-i\g^\nu )(m_{Q'} + \DS p_1 + \DS p_2) \DS \epsilon(p_0) (c_v+c_a\g^5) v(p_3)} {(p_{12} + p_2)^2\Big[(p_1 + p_2 )^2 - m_{Q'}^2 \Big]},
\nn\\
&M_b[n]&= - \kappa \frac{\bar u(p_2) \DS \epsilon(p_0)(c_a\g^5-c_v )(m_b - \DS p_1 - \DS p_3)(-i\g^\nu)\Pi_{p_1}^{[n]} (-i\g^\nu )v(p_3)} {( p_{11} + p_3)^2((p_1 + p_3 )^2 - m_b^2 )},
\nn\\
&M_c[n]&=  - \kappa \frac{\bar u( p_2 )(-i\g^\nu )(m_b + \DS p_{11} + \DS p_2 + \DS p_3 )\DS \epsilon(p_0) (c_a\g^5-c_v) \Pi_{p_1}^{[n]} (-i\g^\nu )v(p_3)} {(p_{11} + p_3)^2 \Big[(p_{11} + p_2 + p_3)^2 - m_b^2 \Big]},
\nn\\
&M_d[n]&= - \kappa \frac{{\bar u( {{p_{2}}} )(-i\g^\nu )\Pi_{p_1}^{[n]} \DS \epsilon(p_0) (c_v + c_a\g^5)(m_{Q'} - \DS p_{12} - \DS p_2 - \DS p_3} )(-i\g^\nu )v(p_3)} {(p_{12} + p_2 )^2 \Big[(p_{12} + p_2 + p_3 )^2 - m_{Q'}^2\Big]}.
\label{eq:3}
\eeq
\ewt
In which, the spin projector $\Pi_{p_1}^{[n]}$ can be written as~\cite{Bodwin:2002cfe}
\beq
&&\Pi_{p_1}^{[^1S_0]} = \frac1{\sqrt {2M_{bQ'}}} \g^5(\DS p_1 + M_{bQ'})
\nn\\
&&\Pi_{p_1}^{[^3S_1]} = \frac1{\sqrt {2M_{bQ'}}} \DS \varepsilon (\DS p_1 + M_{bQ'}).
\label{eq:8}
\eeq
Meanwhile, in order to keep gauge invariance, we adopt $M_{bQ'} \simeq m_b + m_{Q'}$. And the color factor $\Cijk $ can be easily obtained from Fig.~\ref{fig:1} which have the following form:

\beq
\Cijk = N \sum\limits_{a,m,n} \Ta{im}\Ta{jn} \Gmnk,
\label{eq:10}
\eeq
here $k$ stand for the color indices of the diquark and $a = (1, \cdots,8)$ is the incoming gluon.  $N = \sqrt {1/2}$ is the normalization factor. And $i, j, m, n = (1, 2, 3)$ indicate color indices of the two outgoing negative quarks and the two constituent active quarks in the diquark state, respectively. For the $\bar {\bf3}(\bf 6)$ state, the function $G_{mnk}$ is identical to the function $\varepsilon_{mjk}(f_{mjk})$. The antisymmetric function $\varepsilon_{mjk}$ and symmetric function $f_{mjk}$ obeys
\beq
\varepsilon_{mjk}\varepsilon _{m'j'k} = \dd_{mm'}\dd_{jj'}- \dd_{mj'}\dd_{jm'}
\nn\\[2ex]
f_{mjk}f_{m'j'k} = \dd_{mm'}\dd_{jj'} + \dd_{mj'}\dd_{jm'}
\label{eq:11}
\eeq
For the color $\mathbf {\bar 3}$ and $\mathbf {6}$ diquark production, the $\Cijk^2 = 4/3$ and $\Cijk^2 = 2/3$, respectively.

	  \SecIII

Before we calculate the numerical results, we first presenting the choices of the input parameters. $1.8~{\rm GeV}$ and $5.1~{\rm GeV}$ are the masses of $c$ and $b$-quark, respectively. The mass of $Z$-boson was given by PDG~\cite{ParticleDataGroup:2018ovx} with $m_Z=91.1876~{\rm GeV}$. And the value of $|\Psi_{bc}(0) |^2(|\Psi_{bb}(0)|^2)$, we adopted $0.065~{\rm GeV^3}( 0.152~{\rm GeV^3})$, which are consistent with Ref.~\cite{Baranov:1995rc}. For the mass of $\Xc$ and $\Xb$ baryon are taken as $m_{\Xc}=6.9 ~{\rm GeV}$ and $m_{\Xb}=10.2 ~{\rm GeV}$, respectively. The rest of the input parameters as the following numerical values ~\cite{ParticleDataGroup:2018ovx}: $G_F=1.1663787\times 10^{-5}$ denotes the Fermi constant and the Weinberg angle $\theta_w={\rm arcsin}\sqrt {0.2312}$; we utilize $2m_c(2m_b)$ as the renormalization scale $\mu_r$ for the indirect production of $\Xc(\Xb)$.

The decay widths of two main $Z$-boson decay channels for the production of $\XQ$ are demonstrated in Table~\ref{Table:mc1}. Inspecting Table~\ref{Table:mc1}, one can see that, for the production of the $\Xb$, the state of $\nSn{3}{1}_{\bar 3}$ plays the leading role, which the contribution from the state of the $\nSn{3}{1}_{\bar 3}$ can reach to twice than that of $\nSn{1}{0}_6$. As for the $\Xc$ productions, the situations become just the analogical. Moreover, in the case of $\Xc$, the contributions from the decay channel $Z\to c\bar c $ is very small comparable to $Z\to b\bar b$ channel, which is only a few percent of $Z\to b\bar b$ channel.

\btab[t]
\caption{The predicted decay widths $\G(Z\to Q\bar Q)$ (in unit: $10^{-6}$ GeV) with $Q = (c,b)$ for $\Xc$ and $\Xb$ baryon from each $Z$-boson decay channel.}
\btabu{l c c c c c c c c c c c}
\hline
\multirow{2}*{$\G(Z\to Q\bar Q)$} &~& \multicolumn{4}{c}{$\Xc$}   &~& \multicolumn{2}{c}{$\Xb$}
\\ \cline{3-6} \cline{8-9}
		  &&\,$\nSn{3}{1}_{\bar 3}$\, &\,$\nSn{3}{1}_6$\,&\,$\nSn{1}{0}_{\bar 3}$\,&\,$\nSn{1}{0}_6$\,&&\,$\nSn{3}{1}_{\bar 3}$\,&\,$\nSn{1}{0}_6$\,\\ \hline
$Z\to c\bar c $        && 0.644             & 0.322       &0.741            & 0.371        && -    &- \\
$Z\to b\bar b $       && 33.01           & 16.51      & 24.14          & 12.07        && 1.999 &1.028 \\ \hline
\etabu
\label{Table:mc1}
\etab

\btab[b]
\center
\caption{Predicted decay widths (in unit: GeV), branching fraction and events of $\Xc$ and $\Xb$ baryon in $Z$-boson decay.}
\btabu{l l l}
\hline
~~~~~~~~~~~~~~~~~~~~~~~~~~~~~~&  $Z \to \Xc$~~~~~~~~~~~~~~~~~&  $Z\to \Xb$                    \\ \hline
$\G({Z \to \XQ})$          &  $89.71\times 10^{-6}$     &  $3.027\times 10^{-6}$    \\
$\B({Z \to \XQ})$        &  $35.95\times 10^{ - 6}$             &  $1.213\times {10^{ - 6}}$          \\
LHC events        &  $35.95 \times {10^3}$               &  $1.213 \times {10^3}$              \\
CEPC events       &  $35.95 \times {10^6}$               &  $1.213 \times {10^6}$              \\ \hline
\etabu
\label{tab:42}
\etab

In order to assess the doubly heavy baryon $\XQ$ events generated at the LHC(CEPC), the corresponding branching ratio needs to be obtained from the total decay width of the $Z$-boson. Here the total decay width of the $Z$-boson is considered to be $2.495~{\rm GeV}$, which is consistent with Ref.~\cite{ParticleDataGroup:2018ovx}. At the LHC(CEPC), there are about $10^9(10^{12}) $ $Z$-bosons can be produced per year~\cite{LHCLCStudyGroup:2004iyd,CEPCStudyGroup:2018ghi}. According to these conditions mentioned above, the produced events of the double heavy baryon $\XQ$ can be predicted at the LHC(CEPC). We listed the predicted total decay widths, branching ratio and events of $\Xc$ and $\Xb$ baryon via $Z$-boson decay in Table~\ref{tab:42}, where the contribution from each diquark state of $Z$-boson decay channel has been taken into account in total decay widths. From the Table~\ref{tab:42}, we can get the following conclusions
\bit
\ii For production of $\Xb$ baryon, the branching ratio $\B(Z\to\Xb+X)$ is about $10^{-6}$, which is comparable with the results given in Ref.~\cite{Ali:2018ifm}.
\ii Branching ratio of $\B(Z\to\Xc+X)$ amounts to $10^{-5}$ for the production of $\Xc$ baryon, which is comparable with the predictions of $\B(Z\to B_c+X)$~\cite{Deng:2010aq}.
\ii At the CEPC, there are about $10^7(10^{6})$ $\Xc(\Xb)$ baryon can be obtained per year.
\ii Compared to CEPC, there are only about $10^4(10^3)$ $\Xc(\Xb)$ events produced at the LHC, but the upgrades program of HE(L)-LHC will improving the $Z$-boson yield events to a large extent, thus there would produce more $\XQ$ events.
\ii We utilize decay chains of $\Xi^+_{bc}\to\Xi^{++}_{cc}+X\simeq 7\%$~\cite{Qin:2021wyh}, $\Xi^{++}_{cc}\to \Lambda^+_c K^-\pi^+\pi^+\simeq 10\%$~\cite{Yu:2017zst} and $\Lambda^+_c\to pK^+\pi^+\simeq5\%$~\cite{LHCb:2013hvt}, there will about $10^4$ reconstructed $\Xi^+_{bc}$ events can be collected at CEPC, which is comparable to $\Xi^{++(+)}_{cc}$~\cite{Luo:2022jxq}, indicating the observability of the $\Xi^{+}_{bc}$ baryon via $Z$-boson decay.
\eit
\bfig[t]
\grap{0.42}{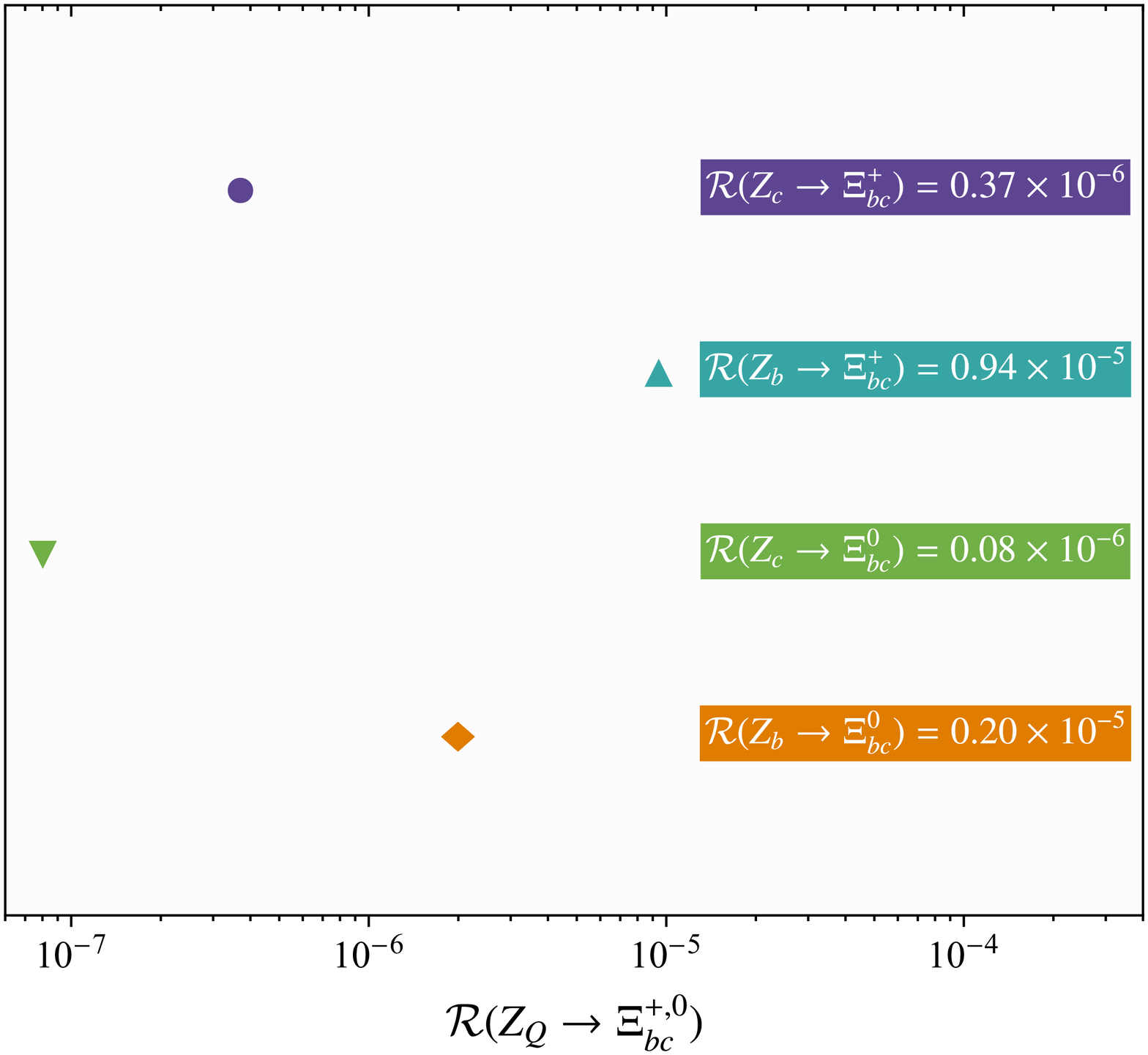}
\caption{Predictions for $\R(Z_Q \to \Xi_{bc}^{+,0})$ with four different $Z$ decay channels. In which, the renormalization scale is setting near $\mu_r = 2m_c$.}
\label{Fig:1}
\efig
Furthermore, in order to predicts the ratio for the production rate of $\Xi^{+,0}_{bc}$ in $Z$-boson decay to $\Lambda^+_c$ accompanied with $K^-\pi^+\pi^+$, e.g. $\R(Z_Q\to\Xi^{+,0}_{bc})$, one can take which have the following expression
\beq
\R(Z_Q\to\Xi^{+,0}_{bc})&=&\frac{\G(Z_Q \to\Xi^+_{bc})}{\G(Z_Q \to\Lambda^+_c)}\times\B(\Xi^{+,0}_{bc}\to\Xi^{++}_{cc})
\nn\\
&\times&\B(\Xi^{++}_{cc}\to \Lambda^+_c K^-\pi^+\pi^+).
\eeq
Here we use the abbreviation $Z_Q$ denote the decay channel $Z\to Q\bar Q$ with $Q = (c,b)$ for convenience. Firstly, due to the total decay width can be related to the brancing fraction directly, one can use the formula $\B({Z_Q \to\Lambda^+_c})=\B(Z_Q)\times f(Q\to\Lambda^+_c)$ to obtain the $\G(Z_Q \to\Lambda^+_c)$. From the PDG, we have $\B(Z_c)=0.12$, $\B(Z_b)=0.15$~\cite{ParticleDataGroup:2020ssz}. The fragmentation fractions of heavy quark to a particular charmed hadron $f(c\to\Lambda^+_c)=0.57$, $f(b\to\Lambda^+_c)=0.73$ are taken from Ref.~\cite{Gladilin:2014tba}. Then, we have
\beq
&&\B(Z_c\to\Lambda^+_c)=6.84\times 10^{-3},\nn\\
&&\B(Z_b\to\Lambda^+_c)=10.95\times 10^{-3}.
\eeq
Secondly, the decay widths of each $Z$-boson decay processes e.g., $Z_Q\to\XQ+X$ have been calculated in this paper, which are listed in Table~\ref{Table:mc1}.  According to the decay chains of $\Xi^{+,0}_{bc}\to\Xi^{++}_{cc}+X\simeq 7\%(1.5\%)$~\cite{Qin:2021wyh}, $\Xi^{++}_{cc}\to \Lambda^+_c K^-\pi^+\pi^+\simeq 10\%$~\cite{Yu:2017zst} and $\Lambda^+_c\to pK^+\pi^+\simeq 5\%$~\cite{LHCb:2013hvt}, we can get the final results shown in Fig.~\ref{Fig:1}. In which the renormalization scale $\mu_r$ is set to be $2m_c$. One can be obtained $\R(Z_b\to\Xi^+_{bc})$ is one magnitude large than $\R(Z_c\to\Xi^+_{bc})$, indicate the decay channel $Z\to b\bar b$ provide key contributions than $Z\to c\bar c$ channel for the indirect production of $\Xi^{+,0}_{bc}$. Comparing $\Xc$ to predicts $\Xi_{cc}$~\cite{Luo:2022jxq} in $Z$ decay, there is a largely gap between $\R(Z_Q\to \Xi^{+,++}_{cc})$ and $\R(Z_Q\to \Xi^{+,0}_{bc})$ about one magnitude, which demonstrates that it is more difficult to collect $\Xi^{+,0}_{bc}$ than $\Xi^{+,++}_{cc}$ at experimental. Moreover, the predicts of $\R(Z_Q\to \Xi^0_{bc})$ via decay channel $\Lambda^+_c\pi^-$ to be $10^{-6}$ order~\cite{Wang:2017mqp}, and our predicts of $\R(Z_Q\to \Xi^0_{bc})$ via decay channel $\Xi^{++}_{cc}$ to be $10^{-5}$ order in $Z$ decay, thus it is more feasible to observe $\Xi^0_{bc}$ through $\Xi^0_{bc}\to\Xi^{++}_{cc}+X$ than via $\Xi^0_{bc}\to\Lambda^+_c\pi^-$ decay channel and also representing its experimental observability.

To further studies for the production of $\XQ$ with $Q' = (b,c)$ via these considered decay channel and usful for experimental research, the $d\G/ds_{ij}$ and the differential decay widths of $\XQ$ with respect to $z$-distributions are plotted in Figs.~\ref{fig:xs1} and~\ref{fig:x1}, we define $s_{i j} = (p_i + p_j)^2$ is the invariant mass and the energy fraction $z = 2E_1/E_Z$, where $E_1$ and $E_{Z}$ are denotes the energy of the $\XQ$ and $Z$-boson, respectively.
\bfigs[t]
\grap{0.49}{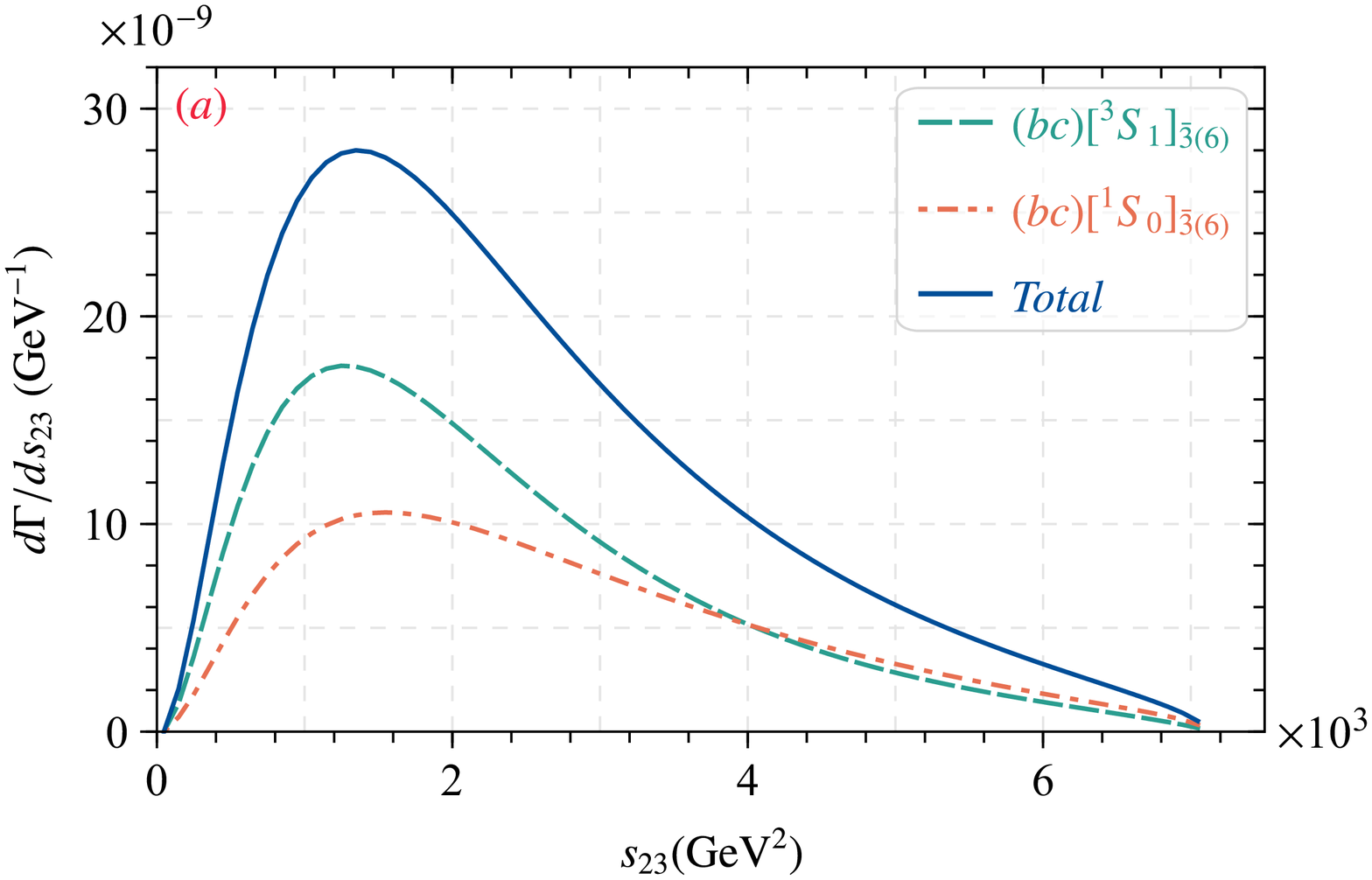}\grap{0.49}{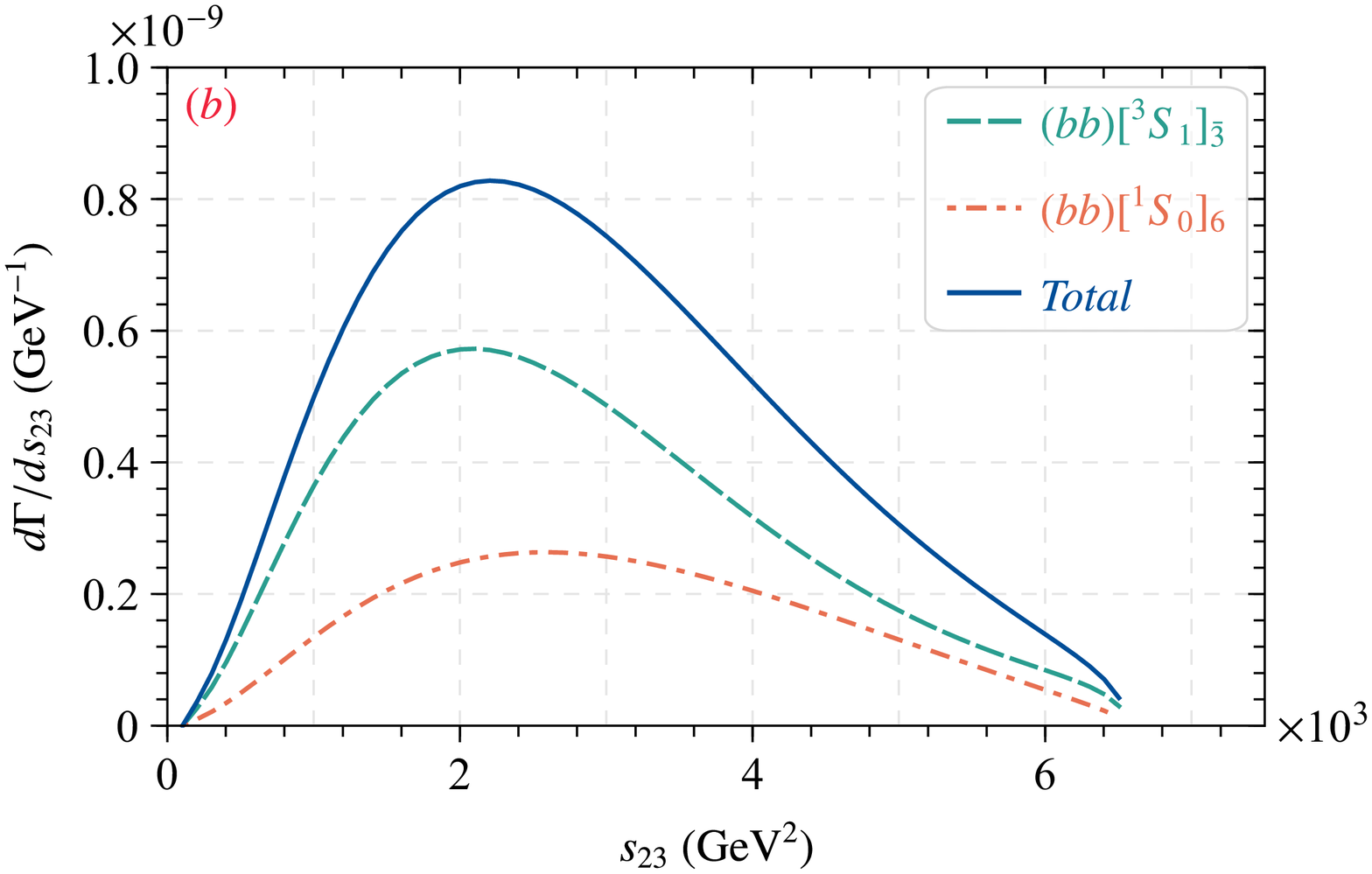}
\caption{The behavior of $d\G/ds_{23}$ for the process $Z \to \Xc(\Xb)+X $, where $\mathbf{\bar 3(6)}$ stands for the color quantum number is the $\bar {\mathbf 3}(\mathbf 6)$ of diquark state ``{\it Total}'' denote the total decay widths that means the each diquark state have been summed.}
\label{fig:xs1}
\efigs
\bfigs[t]
\grap{0.49}{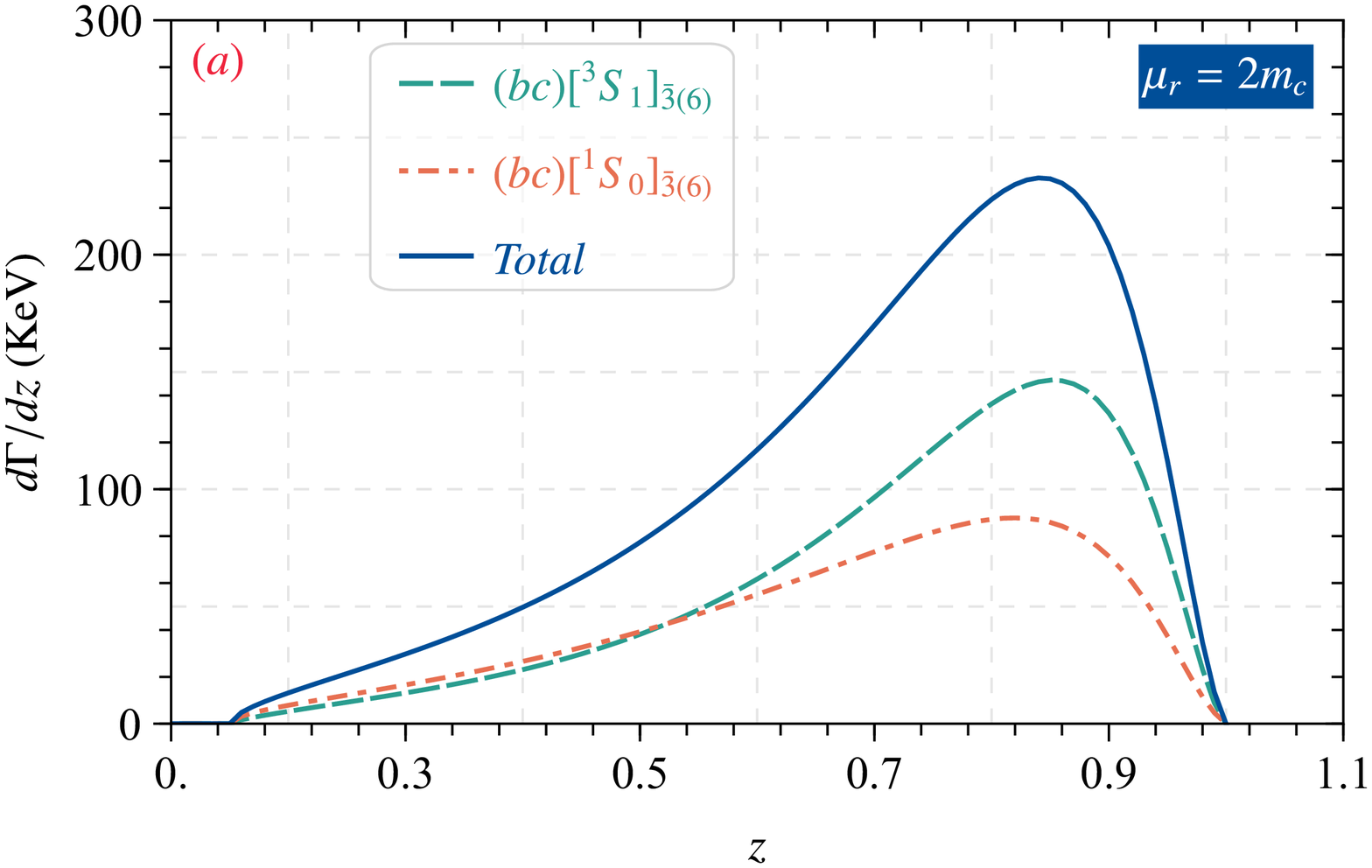}\grap{0.49}{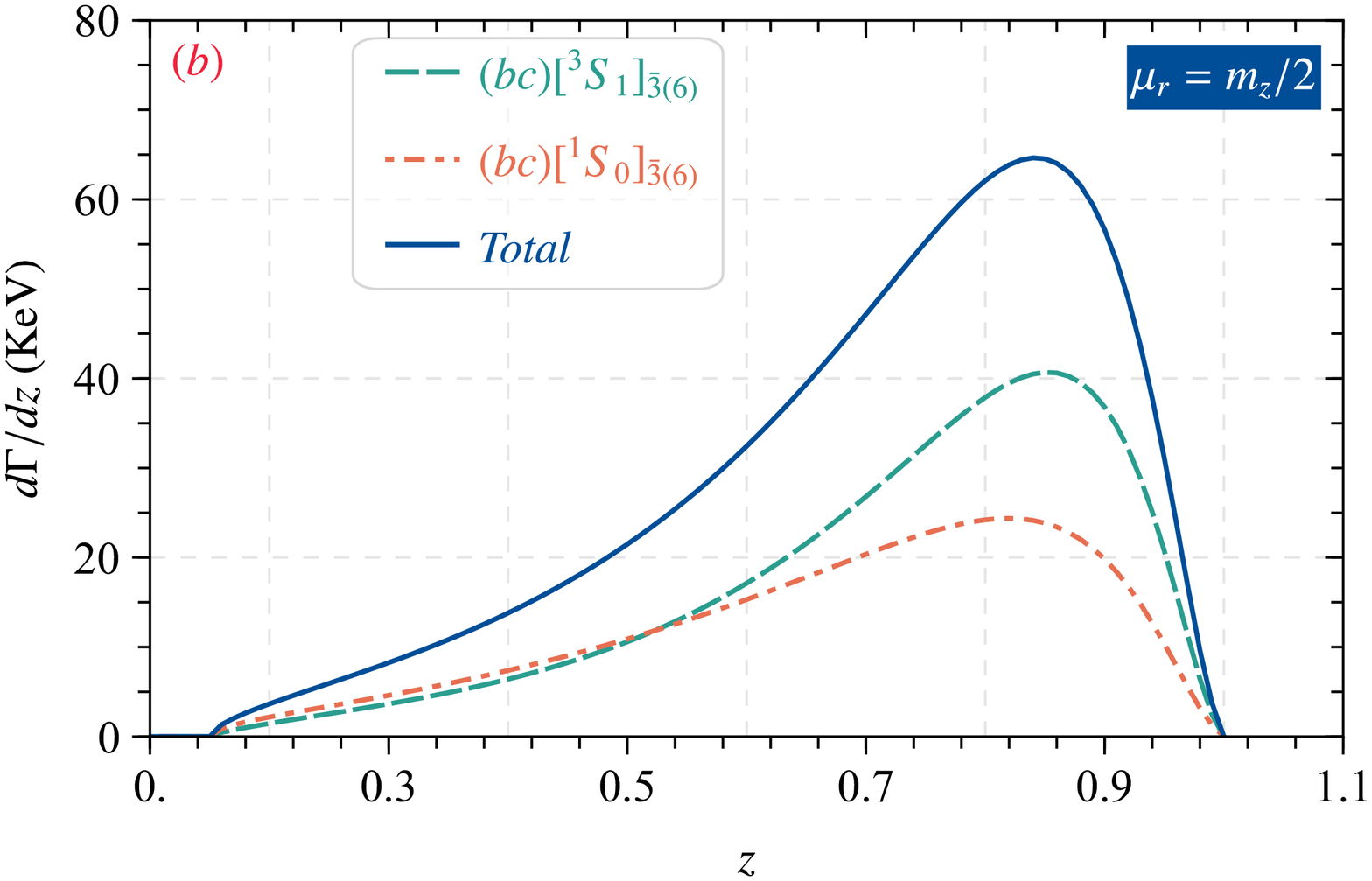}
\grap{0.49}{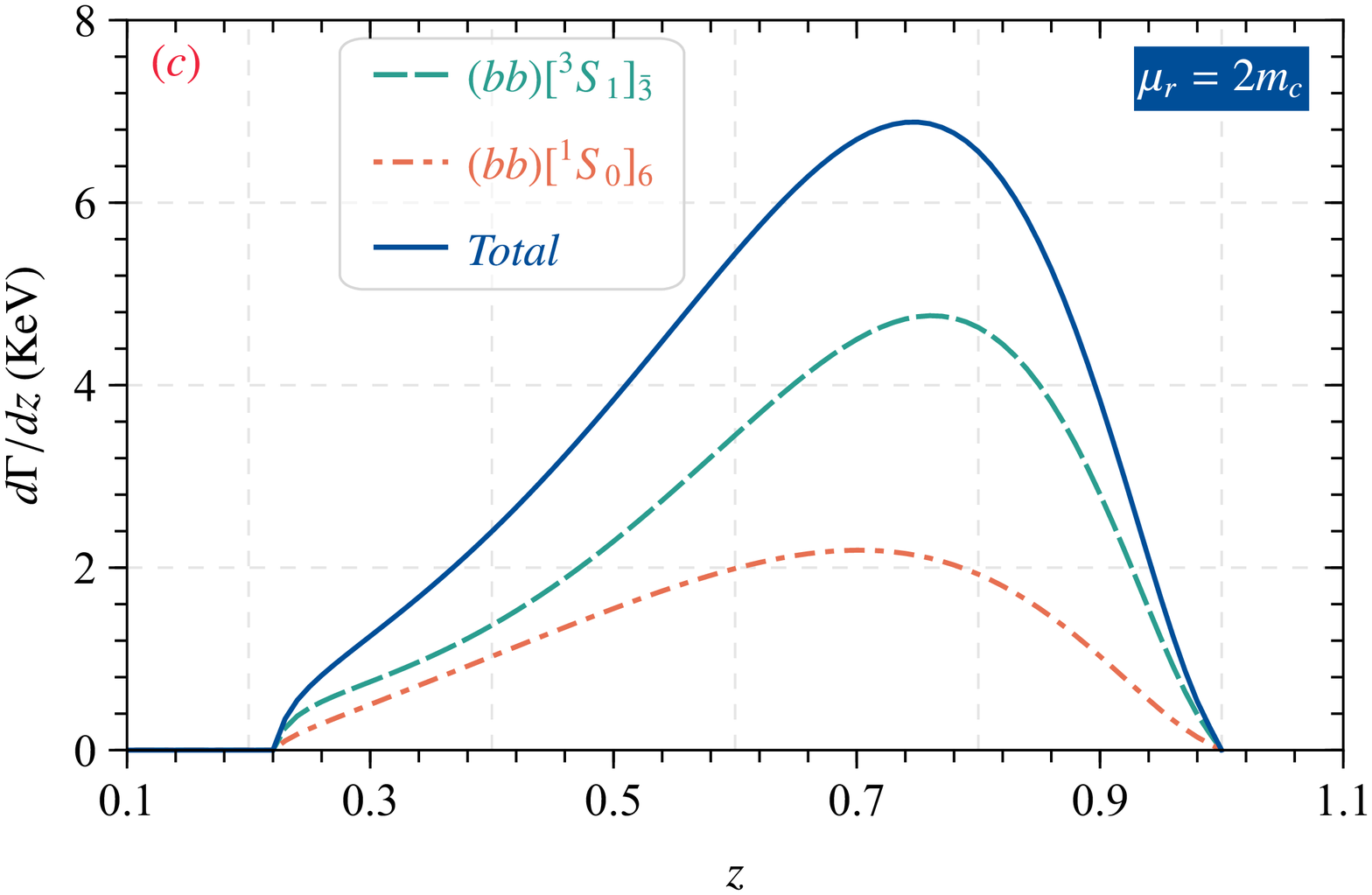}\grap{0.49}{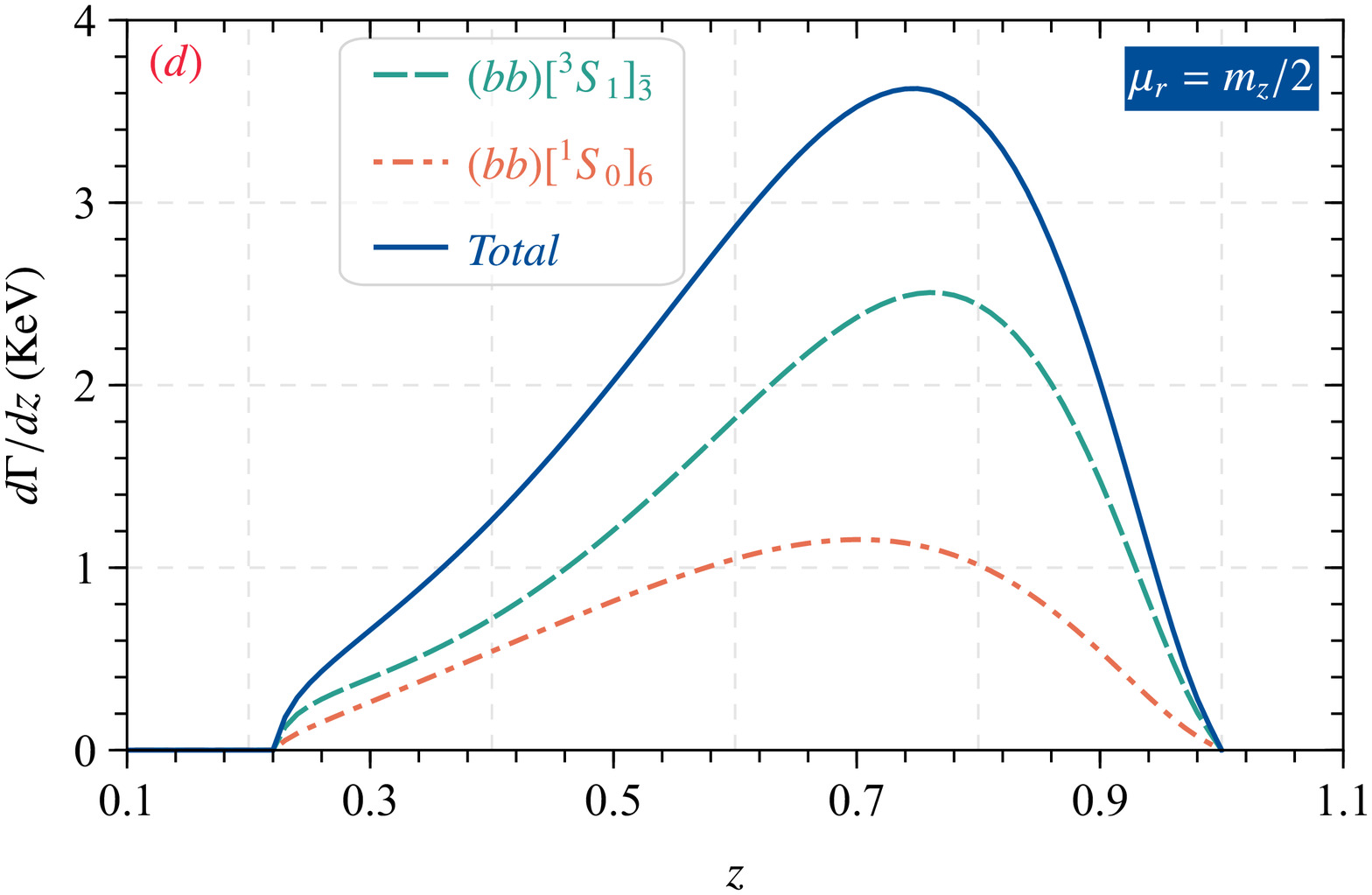}
\caption{The behavior of $d\G/dz$ for the process $Z \to \Xc(\Xb)+X $, where $\mathbf{\bar 3(6)}$ stands for the color quantum number is the $\bar {\mathbf 3}(\mathbf 6)$ of diquark state ``{\it Total}'' denote the total decay widths that means the each diquark state have been summed.}
\label{fig:x1}
\efigs

\btabs
\caption{The decay width $\G$ (in unit: $10^{-6}$) of the process $Z\to\Xc(\Xb)+X$ within theoretical uncertainties through changing the mass of charm quark $m_c = 1.80 \pm 0.05~{\rm GeV}$ and bottom quark $m_b = 5.1\pm 0.5 ~{\rm GeV}$.}
\btabu{ l l l l l l l l l l l l}
\hline
\multirow{2}*{$\mu_r$}~~~~~~~~~~~~~~&\multirow{2}*{$m_Q$}~~~~~~~~~~~~~~~~~~~~~~~~ & &\multicolumn{4}{c}{$\Xc$} &~~~~~~~~&\multicolumn{2}{c}{$\Xb$}&
\\ \cline{4-7} \cline{9-10}
			    & & &$\nSn{3}{1}_{\bar 3}$~~~~~~~~~ &$\nSn{3}{1}_6$~~~~~~~~~ &$\nSn{1}{0}_{\bar 3}$ ~~~~~~~~~&$\nSn{1}{0}_6$& &$\nSn{3}{1}_{\bar 3}$ ~~~~~~~~~&$\nSn{1}{0}_6$\\ \hline
&$m_c = 1.75 ~{\rm GeV} $      & &38.34 & 19.17 &28.08 & 14.04 &&1.999&1.028 \\
$2m_c$&$m_c = 1.80 ~{\rm GeV}$      & & 34.40 & 17.20 &25.41 &12.71 &&1.999&1.028 \\
&$m_c = 1.85 ~{\rm GeV}$      &  & 30.68 & 15.34 &23.05 & 11.53 &&1.999&1.028 \\ \hline
&$m_c = 1.75 ~{\rm GeV} $      &   & 10.47 & 5.236 &7.669 &3.835 &&1.053&0.542 \\
{$m_Z/2$}&$m_c = 1.80 ~{\rm GeV}$  &  & 9.552 & 4.776 &7.057 &3.528 &&1.053&0.542 \\
&$m_c = 1.85 ~{\rm GeV}$      & & 8.736 & 4.368 & 6.510 & 3.255 &&1.053&0.542 \\ \hline
&$m_b = 5.60 ~{\rm GeV} $       & & 34.92 & 17.46 &25.00 & 12.50 &&1.345&0.697 \\
{$2m_c$}&$m_b = 5.10 ~{\rm GeV}$       & & 34.40 & 17.20 &25.41 &12.71 &&1.999&1.028 \\
&$m_b = 4.60 ~{\rm GeV}$      & & 33.89 & 16.95 &25.91 & 12.96 &&3.044&1.385 \\ \hline
&$m_b = 5.60 ~{\rm GeV} $       & & 9.696 & 4.848 &6.943 &3.471 &&0.750&0.389 \\
{$m_Z/2$}&$m_b = 5.10 ~{\rm GeV}$       & & 9.552 & 4.776 &7.057 &3.528 &&1.053&0.542 \\
&$m_b = 4.60 ~{\rm GeV}$       & & 9.412 & 4.706 &7.196 &3.598 &&1.515&0.773 \\ \hline
\etabu
\label{mc1}
\etabs

In Figs.~\ref{fig:xs1}, one can find in cases of $\Xi_{bc}$ and $\Xi_{bb}$ productions, except the state of $[^3S_1]$ plays the leading role, the curves $d\G/ds_{23}$ have a similar variation behavior, which first growing and then dropping with $s_{23}$, and there is a peak in the small region of $s_{23}$.  In Fig.~\ref{fig:x1}, it can be seen that the behavior of the energy fraction $z$ distribution is analogous to that of the invariant mass $s_{ij}$ distribution, which first up and then down with $z$. Precisely, in the cases of $\Xb$ productions, the peak of $\frac{d\G}{dz}| {_{(bb){{{[^3}{S_1}]}_{\bar 3 }}}} $ is around $z = 0.75$ and $\frac{d\G}{dz}|_{(bb)[^1S_0]_6}$ peaks near $z = 0.7$. As for the $\Xc$, the peak of $\frac{d\G}{dz}|_{(bc)\nSn{1}{3}_{\bar 3(6)}} $ is around $z = 0.8$ and $\frac{d\G}{dz}|_{(bc)[^1S_0]_{\bar 3(6)}}$ peaks near $z = 0.85$. Due to the dominant effect of quark fragmentation mechanism, the peak of the $\Xc(\Xb)$ energy distribution in $Z\to \Xi_{bc(bb)}+X$ is located in the larger $z$-region.

Then, to make a discussion about the theoretical uncertainties for the process $Z\to\XQ+X$ precisely, we shall focus the attention on analyze caused the uncertainty from the mass of $c$ and $b$-quark, and the renormalization scale. And the uncertainties from the $|\Psi_{bQ'}(0)|^2 $ does not discussed, since the $|\Psi_{bQ'}(0)|^2 $ is an overall coefficient in the calculation, which can be computed out easily. The contribution of these considered decay channels has been summarized as the total decay width. The resulting $c$ and $b$-quark mass uncertainties by varying $m_c = 1.80 \pm 0.05~{\rm GeV}$ and $m_b = 5.1 \pm 0.5~{\rm GeV}$ in our calculations,respectively. The caused renormalization scale uncertainties by choices the $\mu_r=2m_c(m_z/2)$ for $\Xc$ and $\mu_r=2m_b(m_z/2)$ for $\Xb$, which are listed in Table~\ref{mc1}. One can see that
\bit
\ii  In the case of the indirect production of the $\Xc$ baryon in $Z$ decay, the decay width lessen with the augment of the mass of $c$-quark, which is largely ascribe the suppression of phase space. To our astonishment, due to the affect of the projector in Eq.~\eqref{eq:8}, the decay width increases with the increment of mass of $b$-quark for the indirect production of $\langle cc\rangle \nSn{3}{1}$ state via $Z$ decay, and the decay width of the process $Z\to\langle cc\rangle\nSn{1}{0}+X$ decreases with the elevate of mass of $b$-quark. Moreover, the uncertainty caused by $m_c$ is relatively larger than those of $m_b$.
\ii For the process $Z\to\Xb+X$ baryon, the decay width $\G$ reduce with the elevate of mass of $b$-quark, both $\Xb[^3S_1]_{\bar 3}$ and $\Xb[^1S_0]_{6}$.
\eit

	  \SecIV

In this work, we complete the studies of the indirect production of $\XQ$ with $Q' = (c, b) $via $Z$-boson decay bases on the framework of NRQCD. By including the contributions of the intermediate diquark states, i.e. $\bQ{c}\nSn{3}{1}_{\bar 3/6}$, $\bQ{c}\nSn{1}{0}_{\bar 3/6}$, $\bQ{b}\nSn{1}{0}_{6}$ and $\bQ{b}\nSn{3}{1}_{\bar 3}$, the branching ratio of $\B(Z\to\Xc+X)$ is about $10^{-5}$, and $\B(Z\to\Xb+X)$ amounts to $10^{-6}$, following which as amount as $10^4(10^7)$ $\Xc$ events and  $10^3(10^6)$ $\Xb$ events produced at the LHC(CEPC). To be beneficial as regards experimental observation, the differential decay widths of $\XQ$ with respect to $z$ distributions have been presented. Moreover, we have estimated the production ratio $\R(Z_Q\to \Xi^{+,0}_{bc})$ of $\Xi^{+,0}_{bc}$ to $\Lambda^+_c$  by $Z$-boson decay channel $c\bar c$ and $b\bar b$ for the first time, the $\R(Z_Q\to \Xi^{+,0}_{bc})$ up to $10^{-6}$ for $c\bar c$ channel and $10^{-5}$ for $b\bar b$ channel, respectively. Abundant $\XQ$ baryon events, considerable branching ratio and $\R(Z_Q\to \Xi^{+,0}_{bc})$, which demonstrate the observability of the $\XQ$ baryon in $Z$-boson decay at the experiment. Thus, we think that it is worthwhile and feasible to hunt $\XQ$ baryon through $Z$-boson decay at the LHC and CEPC.

\acknowledgments
We are grateful for the Professor Zhan Sun's valuable comments and suggestions. This work is supported in part by the Natural Science Foundation of China under Grant No. 12265010, and by the Project of Guizhou Provincial Department of Education under Grant No.KY[2021]030.

\edoc